\documentclass[prb,aps,twocolumn,showpacs,preprintnumbers]{revtex4}
\usepackage{epsfig,epsf,psfrag}
\usepackage{graphicx}
\usepackage{epic,eepic}
\usepackage{color,pstcol}
\begin{document}
\title{Detecting entangled states in graphene via crossed Andreev reflection}
\date{\today}

\author{Colin Benjamin and Jiannis K. Pachos}

\affiliation{Quantum Information Group, School of Physics and
Astronomy, University of Leeds, Woodhouse Lane, Leeds LS2 9JT,
UK.}

\begin{abstract}

  Shot noise cross-correlations across single layer graphene
  structures are calculated with insulators separating a
  superconducting region. A new feature of specular crossed Andreev
  reflection comes into play due to the unique band structure of
  graphene. This gives rise to a rich structure in the states of the
  electric current flowing across the graphene sheet. We identified a
  parametric regime where {\em positive} shot noise cross-correlations
  of the current appear signifying entanglement. In contrast to
  previous proposals the sign of the cross-correlations can be easily
  tuned by the application of a gate voltage.

\end{abstract}


\maketitle
\section{Introduction}
Shot noise is defined as the temporal fluctuation of electric current
in a non-equilibrium set-up~\cite{Martin,ref-on-noise}.  When the shot
noise cross-correlations between two regions turn positive it signals
the presence of electronic entangled states~\cite{pt-noise}. In order
to make use of these correlations for quantum information purposes one
would need to spatially separate the electrons without destroying the
entanglement~\cite{melin-cb,buttiker}. This is ideally detected by
entangled electrons traversing different wires~\cite{loss-entangle}.
The quantum correlations can be provided by Cooper pairs present in
superconductors, which is the most entangled state found in nature.

To intuitively understand how shot noise contains the signature of
entanglement we resort to statistics. Shot noise cross-correlations
are defined as cross-correlations of current fluctuations across two
distinct regions. Absence of entanglement leads to positive
correlations for photons (bunching) and negative for electrons
(anti-bunching). The observation of positive shot noise correlations
for electrons is a signature that they are in an entangled state. This
has been most famously predicted in normal metal-superconductor-normal
metal structures~\cite{melin-cb,ref-on-noise,Martin}, but it has not
yet been experimentally demonstrated. An earlier experimental
attempt~\cite{kontos} in a two dimensional electron gas beam splitter
connected to a superconductor could not arrive at any definite
conclusion possibly due to the low tunability of these devices. In
this work we investigate what happens to the noise cross-correlations
when a single layer of graphene replaces the normal metal or
semiconductor. Our motivation comes from the following fact. In
contrast to a normal metal, the energy of transported electrons can be
very efficiently controlled in a graphene layer via the application of
a gate voltage thus being much more amenable to experiments. This was
demonstrated in Ref.~\cite{heersche} where it was shown that the
Josephson current could be very efficiently tuned via the application
of a small gate voltage. We expect that this characteristic of
graphene structures will facilitate the observation of entanglement in
solid-state systems, thus, opening the way for their wider use in
quantum information applications~\cite{teleport}.

\begin{figure}[h]
\centerline{
\includegraphics[width=\linewidth]{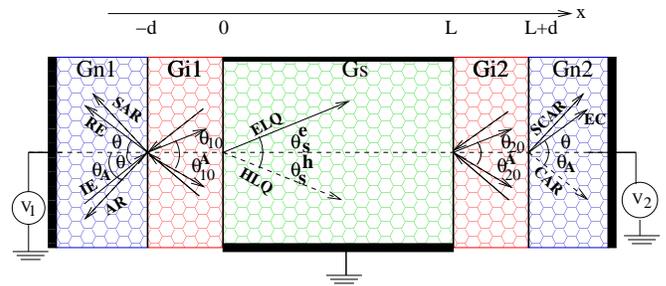}}
\caption{An overview of the setting from the top. Two insulating layers of
graphene (Gi's)  on either side of the superconducting graphene
layer (Gs). Voltages V$_1$ and V$_2$ are applied to either ends of
the normal graphene layers (Gn's). Schematic of specular crossed
Andreev reflection is also depicted. Incident electron at
angle $\theta$ (IE). Reflected electron at angle $-\theta$ (RE). 
Andreev reflected hole at angle $\theta_A$ (AR). Specular Andreev
reflected hole at angle $-\theta_A$ (SAR). Electron like
quasiparticle (ELQ). Hole like quasiparticle (HLQ). Crossed Andreev
reflection at angle $\theta_A$ (CAR). Specular crossed Andreev
reflected hole at angle $-\theta_A$ (SCAR). Electron co-tunnelling at
angle $\theta$ (EC). \label{scheme}}
\end{figure}

\begin{figure}[h]
\centerline{\includegraphics[width=\linewidth]{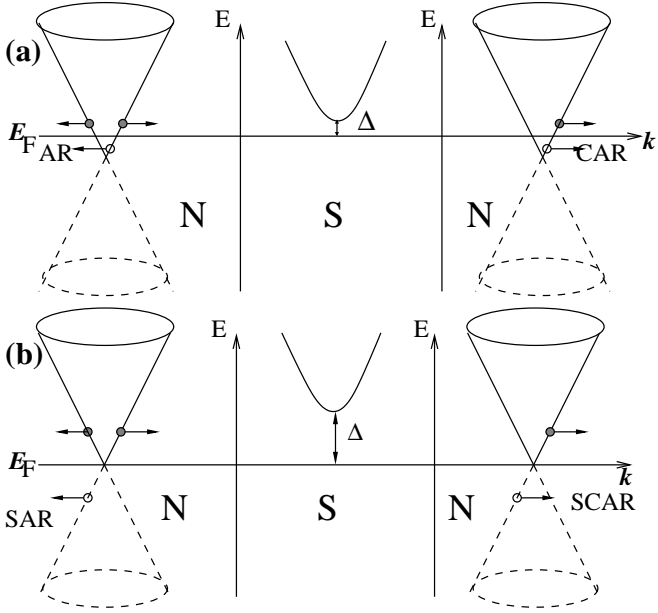}}
\caption{ Energy-Momentum diagram to explain specular crossed Andreev
  reflection where the regions of normal (N) graphene and
  superconducting (S) graphene are as indicated. (a) $E_{F} > \Delta$
  regime where Andreev and crossed Andreev reflection occur in the
  same band, and (b) $ E_{F} \ll \Delta$ regime where Andreev and
  crossed Andreev reflection occur in a specular fashion across the
  bands. If, $E  > E_{F} > \Delta$, where $E$ is the energy of the particle measured from the Fermi level, then also specular Andreev reflection occurs. In all calculations we are in the regime, where $E < \Delta$. \label{energy}}
\end{figure}

\section{Model}
Graphene is a monatomic layer of graphite with a honeycomb lattice
structure~\cite{graphene-rmp} that can be split into two triangular
sublattices $A$ and $B$. The electronic properties of graphene are
effectively described by the Dirac equation~\cite{graphene-sudbo}. The
presence of isolated Fermi points, $K_+$ and $K_-$, in its spectrum,
gives rise to two distinctive valleys. In this work we deal with a
normal-insulator-superconductor-insulator-normal (NISIN) graphene
junction. We consider a sheet of graphene on the $x$-$y$ plane.
Superconductivity is induced via the proximity effect, where a normal
superconductor at close range on top of the sheet generates the
desired superconducting correlations. In Fig.~\ref{scheme} we sketch
our proposed system. The superconducting region is located between
$0<x<L$, while the insulators are located on its left, $-d<x<0$, and
on its right, $L<x<L+d$. The normal graphene planes are to the
left-end, $x<-d$, and to the right-end, $x>L+d$.

As we will demonstrate in the following there are additional processes
occurring at the normal graphene-superconducting graphene-normal
graphene junctions than those seen at normal
metal-superconductor-normal metal
junctions~\cite{Duhot-Melin}.  These are local
specular Andreev reflection and crossed (non-local) specular Andreev
reflection. Importantly, Andreev reflection in graphene can switch the
valley bands, i.e., conduction to valley, see Fig.~\ref{energy}. This
process is known as specular Andreev
reflection~\cite{been-gra-supercon}, explained in Fig.~\ref{energy}.
In the process of normal Andreev reflection, an incident electron from
the normal metal side is reflected as a hole which retraces the
trajectory of the electron.  In specular Andreev reflection, the
reflected hole follows the trajectory which a normally reflected
electron would have. In this work we see, in addition to this, the
possibility of specular crossed Andreev reflection, where a hole is
reflected at the other lead but in a specular fashion (see
Fig.~\ref{scheme}).

For a quantitative analysis we describe our system by the Dirac
Bogoliubov-de Gennes equation that assumes the
form~\cite{graphene-sudbo}
\begin{equation}
  \left(\begin{array}{cc} \hat{H}-E_{F}\hat{I}& \Delta\hat{I}\\
      \Delta^{\dagger}\hat{I}&
      E_{F}\hat{I}-\hat{T}\hat{H}\hat{T}^{-1}\end{array}\right)\Psi=E\Psi,
\end{equation}
where $E$ is the excitation energy, $\Delta$ is the superconducting
gap of a s-wave superconductor, $\Psi$ is the wavefunction and
$\hat{\cdot}$ represents $4\times4$ matrices. In the above equation
\begin{equation}
  \hat{H}=\left(\begin{array}{cc}H_{+}&0\\0&H_{-}\end{array}\right),\,\,\,
  H_{\pm}=-i \hbar v_{F}(\sigma_{x}\partial_{x}\pm\sigma_{y}\partial_{y})+ U
  \nonumber
\end{equation}
Here $\hbar, v_{F}$ (set equal to unity hence forth) are the Planck's
constant and the energy independent Fermi velocity for graphene, while
the $\sigma$'s denote Pauli matrices that operate on the sublattices
$A$ or $B$. $U$ is the electrostatic potential which can be adjusted
independently via a gate voltage or doping. We assume $U=0$, in the
normal regions, while $U=V_{i}, i=1,2$, in either insulating regions
and $U=-U_{0}$ in the superconductor.  The subscripts of Hamiltonian
$\pm$ refer to the valleys of $K_+$ and $K_-$ in the Brillouin zone.
$T=-\tau_{y}\otimes{\sigma_{y}}C,$ ($C$ being complex conjugation) is
the time reversal operator, with $\tau$ being Pauli matrices that
operate on the $\pm$ space and $\hat I$ is the identity matrix.

Let us consider an incident electron from the normal side of the
junction ($x<-d$) with energy $E$. For a right moving electron with an
incident angle $\theta$ the eigenvector and corresponding momentum
reads
\begin{equation}
\psi^{e}_{+}=[1,e^{i\theta},0,0]^{T}e^{i p^{e}\cos\theta x},\,\,\,
p^{e}=(E+E_{F}).
\end{equation}
A left moving electron is described by the substitution $\theta
\rightarrow \pi -\theta$. If Andreev-reflection takes place, a left
moving hole is generated with energy $E$, angle of reflection
$\theta_{A}$ and its corresponding wave-function is given by
\begin{equation}
\psi^{h}_{-}=[0,0,1,e^{-i\theta_{A}}]^{T}e^{-ip^{h}\cos\theta_{A}
x}, \,\,\,p^{h}=(E-E_{F}).
\end{equation}
The superscript e (h) denotes an electron-like (hole-like) excitation.
Since translational invariance in the $y$-direction holds the
corresponding component of momentum is conserved. This condition
allows for the determination of the Andreev reflection angle
$\theta_A$ through $p^{h}\sin(\theta_{A})=p^{e}\sin(\theta)$. There is
no Andreev reflection and consequently no sub-gap conductance for
angles of incidence above the critical angle
$\theta_{c}=\sin^{-1}(|E-E_{F}|/(E+E_{F}))$. In the insulators,
$-d<x<0$ and $L<x<L+d$, the eigenvector and momentum of a right moving
electron are given by
\begin{equation}
\psi^{e}_{iI+}=[1,e^{i\theta_{i0}},0,0]^{T}e^{i p^{e}_{I}\cos
\theta^{}_{i0} x}, p^{e}_{iI}=(E+E_{F}-V_{i}),
\end{equation}
 with $i=1,2$ while a left moving hole is described by
\begin{equation}
\psi^{h}_{iI-}=[0,0,1,e^{-i\theta^{A}_{i0}}]^{T}e^{-ip^{h}_{iI}
\cos \theta^{A}_{i0} x}, p^{h}_{iI}=(E-E_{F}+V_{i}).
\end{equation}

On the superconducting side of the system, ($0<x<L$), the possible
wavefunctions for transmission of a right-moving quasiparticle
with excitation energy $E>0$ read
\begin{eqnarray}
\Psi^{e}_{S+}&=&[u,ue^{i\theta^{+}},v,v e^{i\theta^{+}}]^{T}
 e^{iq^{e}\cos \theta^{+} x},\nonumber\\
\Psi^{h}_{S-}&=&[v,ve^{i\theta^{-}},u,u e^{i\theta^{-}}]^{T}
 e^{iq^{h}\cos \theta^{-} x}.
\end{eqnarray}
with $q^{e}=(E_{F}+U_{0}+\sqrt{E^{2}-\Delta^{2}})$ and
$q^{h}=(E_{F}+U_{0}-\sqrt{E^{2}-\Delta^{2}})$. In the sub-gap regime
the quasi-particle wave-vectors have a small imaginary component as
$q^{e/h}=E_{F}+U_{0}\pm 1/\xi$, where $\xi=1/\Delta$ is the coherence
length. The coherence factors are given by $u=\sqrt{(1+{\sqrt{1-
      \Delta^{2}/E^{2}}})/2}$,
$v=\sqrt{(1-{\sqrt{1-\Delta^{2}/E^{2}}})/2}$. We have also defined
$\theta^{+}=\theta^{e}_{S},$ $\theta^{-}=\pi-\theta^{h}_{S}$. The
transmission angles $\theta^{\alpha}_{S}$ for the electron-like and
hole-like quasi-particles are given by $q^{\alpha}\sin
\theta^{\alpha}_{S}=p^{e}\sin \theta, \alpha=e,h$. In the following we
limit ourselves to the regime where $U_{0} \gg \Delta$, such that the
mean field conditions for superconductivity are satisfied. The
trajectory of the quasi-particles in the insulating region are defined
by the angles $\theta_{i0}$ and $\theta^{A}_{i0}$. These angles are
related to the injection angles by
\begin{eqnarray}
\sin \theta_{i0}/\sin \theta=(E+E_{F})/(E+E_{F}-V_{i}),\nonumber\\
\sin \theta^{A}_{i0}/\sin \theta=(E+E_{F})/(E-E_{F}+V_{i}).
\label{eq-theta}
\end{eqnarray}
Here, we adopt the thin barrier limit defined as, $\theta_{i0},
\theta^{A}_{i0} \mbox{ and } d \rightarrow 0,$ while
$V_{i}\rightarrow\infty$, such that $p^{e}_{iI}d, p^{h}_{iI}d
\rightarrow \chi_{i}$.  To solve the scattering problem, we match the
wavefunctions at four interfaces: $\psi|_{x=-d}=\psi_{1I}|_{x=-d},$
$\psi_{1I}|_{x=0}=\Psi_{S}|_{x=0},$
$\Psi_{S}|_{x=L}=\psi_{2I}|_{x=L},$ and
$\psi_{2I}|_{x=L+d}=\psi|_{x=L+d},$ where, starting with normal
graphene at left,
$\psi=\psi_{+}^{e}+s^{ee}_{11}\psi_{-}^{e}+s^{eh}_{11}\psi_{-}^{h},$
$\psi_{iI}=p_{i}\psi_{iI+}^{e}+q_{i}\psi_{iI-}^{e}+m_{i}\psi_{iI+}^{h}+n_{i}\psi_{iI-}^{h},
i=1,2,$
$\Psi_{S}=p_{S}\Psi^{e}_{S+}+q_{S}\Psi^{e}_{S-}+m_{S}\Psi^{h}_{S+}+n_{S}\Psi^{h}_{S-},$
and finally for normal graphene at the right,
$\psi=s^{ee}_{12}\psi_{+}^{e}+s^{eh}_{12}\psi_{+}^{h}$. Solving these
equations leads to the amplitude of Andreev reflection $s^{eh}_{11}$,
normal reflection $s^{ee}_{11}$, amplitude of electron co-tunnelling
(EC) $s^{ee}_{12}$, and of crossed Andreev reflection (CAR)
$s^{eh}_{12}$.
\section{Results}
\subsection{Specular crossed Andreev reflection}
\begin{figure}
\centerline{\includegraphics[width=3.5in,height=3.5in]{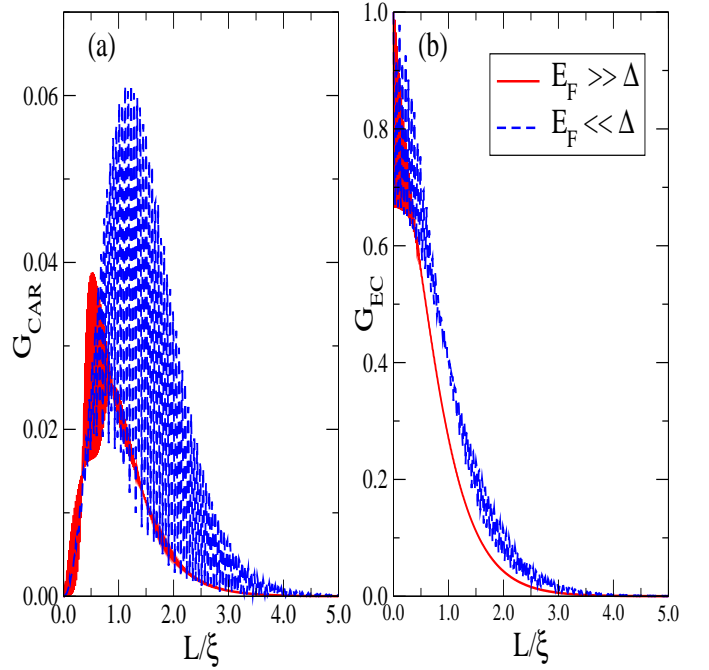}}
\vskip 0.5in
\caption{(a) Crossed Andreev reflection and (b) electronic
co-tunneling as function of the superconducting length, $L/\xi$.
In both figures, $\chi_{1}=-\chi_{2}=\pi/4, U_{0}=1000\Delta$ and
$E=0.15 \Delta$.\label{car-ec}}
\end{figure}

\begin{figure}
\centerline{\includegraphics[width=3.5in,height=3.5in]{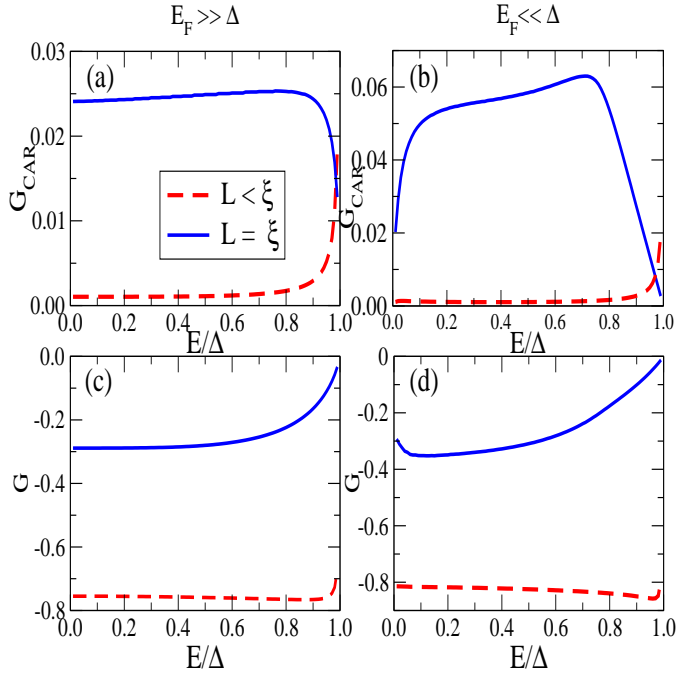}}
\vskip 0.5in
\caption{(a) Crossed Andreev reflection and (b) specular crossed
Andreev reflection. Non-local conductance for a NISIN
graphene based structure as function of the electronic energy for
(c) normal and (d) specular reflection cases. In all figures,
$\chi_{1}=-\chi_{2}=\pi/4,$ and $U_{0}=1000\Delta.$ \label{car-scar}}
\end{figure}

The first issue we tackle is the non-local conductance. Similar caluclations, but for bipolar structures, were performed in Ref.~[\onlinecite{cayssol}].
It is defined as the conductance in the right lead when both
superconduction region and right graphene layer are grounded, while a
voltage is applied to the left graphene sheet. The non-local
conductance is given by the difference between the crossed Andreev and
electronic co-tunneling currents in the absence of a bias at right,
where $G=G_{CAR}-G_{EC},$ with~\cite{Duhot-Melin}
\begin{equation}
 G_{CAR}=\int_{-\frac{\pi}{2}}^{\frac{\pi}{2}}\!\!\! d\theta
\cos\theta_{A} |s^{eh}_{12}|^{2},
G_{EC}=\int_{-\frac{\pi}{2}}^{\frac{\pi}{2}}\!\!\! d\theta \cos\theta
|s^{ee}_{12}|^{2}.
\end{equation}
In the following figures all the quantities are in their dimensionless
form with the superconducting gap set to $\Delta=1$.  The other energy
parameters are expressed in terms of $\Delta$. In Fig.~\ref{car-ec} we
plot the non-local CAR and EC current as function of the length of the
superconducting region. We differentiate between two regimes.  First,
for $E_{F} \gg \Delta$ there is absence of interband non-local
electron-hole transmission, denoted as the crossed Andreev regime
(Fig.~\ref{energy}). Second, for $E_{F} \ll \Delta$ non-local
interband electron-hole transmission is permitted giving rise to the
specular crossed Andreev regime. As function of the length we find
that both non-local coefficients vanish for large values. However,
while the EC current decreases almost monotonically from a peak at
$L\ll \xi$ to vanishing for $L \gg \xi$, the CAR current is maximum
around $L \sim \xi$, and it vanishes for the extreme cases $L \ll
\xi$, $L \gg \xi$. In Fig.~\ref{car-scar}, we plot the crossed Andreev
current for normal transmission (left) as well as specular reflection
(right). We observe that the specular CAR current might dominate the
normal current in the $E \ll \Delta$ regime. One very interesting
fact, which is partly seen in NS graphene junctions, is that, just
like the specular Andreev reflection seen there, here too the crossed
specular Andreev reflection is reduced to vanishing at $E \sim
\Delta$, but the normal crossed Andreev current is marginally reduced
at $E \sim \Delta$. However, the non-local conductance (see
Fig.~\ref{car-scar}), is dominated by electron co-tunnelling.  It is
also periodic as function of the strength of the insulating barrier's
$\chi_{i}$'s (not plotted here)~\cite{sengupt-gra}.

\begin{figure}
\centerline{\includegraphics[width=3.5 in, height=4.0in]{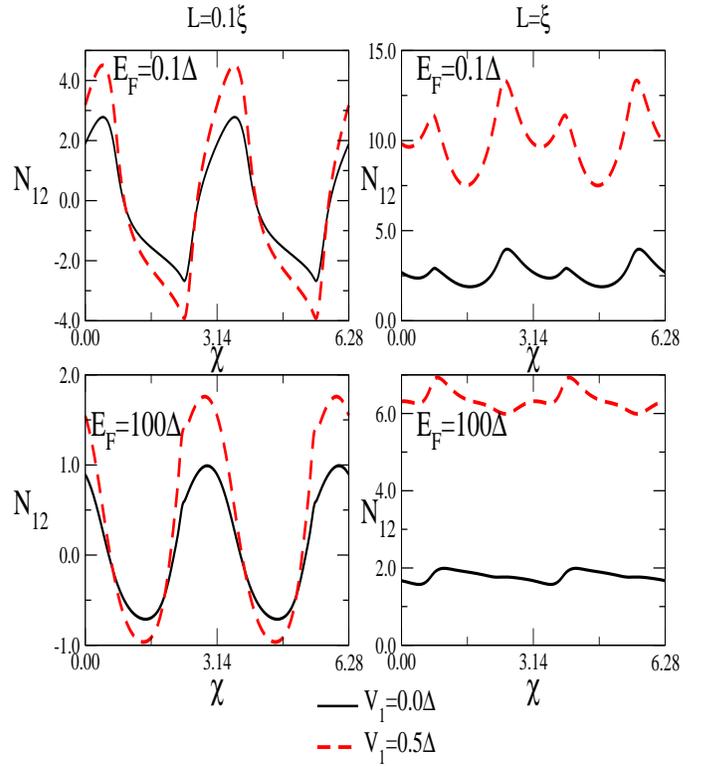}}
\vskip 0.5in
\caption{Noise cross-correlations as function of the gate voltage
($\chi=\chi_1$) applied to the left insulator. The right insulator
is fixed at gate voltage $\chi_{2}=0$, while $U_{0}=1000\Delta$ and $V_{2}=0.2\Delta$. }
\label{noise1}
\end{figure}
\subsection{Shot noise cross-correlations}
Next we calculate the shot noise cross-correlations, which is the main
focus of our work. For that we first have to derive an expression for
the shot noise in multi-terminal settings~\cite{ref-on-noise} applied
to graphene. The fluctuations of the current away from the average is
termed noise. A general expression for current fluctuations between
any two arbitrary leads is given by
\begin{equation}
N_{ij}(\tau)=\langle\Delta \hat{I_{i}}(t)\Delta
\hat{I_{j}}(t+\tau)+\Delta \hat{I_{j}}(t+\tau)\Delta
\hat{I_{j}}(t)\rangle, \label{eq:noise-def}
\end{equation}
where $\Delta
\hat{I_{i}}(t)=\hat{I_{i}}(t)-\langle\hat{I_{i}}(t)\rangle$. The
Fourier transform of Eq.~(\ref{eq:noise-def}) gives
\begin{equation}
N_{ij}(w)\delta(w+w')=\langle\Delta \hat{I}_{i}(w')\Delta
\hat{I}_{j}(w)+\hat{I}_{j}(w)\Delta \hat{I}_{i}(w')\rangle.
\label{eq:noise-2}
\end{equation}
For simplicity we consider the experimentally feasible zero
frequency noise limit, where displacement currents are absent. The
current operator is given by
\begin{equation}
\hat{I}_{i}(w=0)=\sum_{\scriptsize \begin{array}{c}k,l \in Gn_{1},Gn_{2},Gs\\
\alpha,\beta,\gamma,\delta \in  e,h\end{array}} q_{\alpha}\int dE
A_{k\gamma;l\delta}(i,\alpha)
\hat{a}^{\dagger}_{k\gamma}\hat{a}_{l\delta}, \label{eq:curr}
\end{equation}
with $A_{k\gamma,l\delta}(i,\alpha)=\delta_{ik}\delta_{il}
\delta_{\alpha\gamma}\delta_{\alpha\delta}-
s^{\alpha\gamma\dagger}_{ik}s^{\alpha\delta}_{il}$, where Greek
indices denote the nature ($e$ for electrons, $h$ for holes) of
the incoming/outgoing particles with their associated charges
$q_\alpha$, while Latin indices $l$, $k$ identify the graphene
sheets and $\hat{a}_{l\delta}$ denotes annihilation operator for a
particle in lead $l$ with charge $\delta$. From Eqs.
(\ref{eq:noise-2}) and (\ref{eq:curr}) the zero frequency noise
cross-correlations between the currents at left and right normal
graphene sheets ($Gn_{1},Gn_{2}$) become~\cite{ref-on-noise}
\begin{eqnarray}
\label{noise_cross}
N_{12}\!\!\!&=&\!\!\!\!\sum_{\scriptsize \begin{array}{c}k,l \in Gn_{1},Gn_{2},Gs\\
    \alpha,\beta,\gamma,\delta \in  e,h\end{array}}\!\!\!\!\!\!\!\!
\frac{q_{\alpha}q_{\beta}}{h}\int_{-\frac{\pi}{2}}^{\frac{\pi}{2}}{d\theta}
\cos \theta \int dE A_{k\gamma,l\delta}(1,\alpha)\nonumber\\
&& \,\,\,\,\,\,\,\,\,\,\,\,\,\,\,\,\,\,\,\,\,\,\times
A_{l\delta,k\gamma}(2,\beta)f_{k\gamma}(1-f_{l\delta})
\end{eqnarray}
 $f_{k\gamma}$ is a Fermi function for particles of type $\gamma$ in graphene sheet $k$.

 In the limit $L \ll \xi$ Andreev and cross-Andreev reflection vanish,
 which implies that in this limit noise-correlations are
 negative~\cite{melin-cb}. In the limit $L \gg \xi$ both non-local
 currents vanish leading to vanishing noise cross-correlations.
 However, it is the length in-between these limits where noise not
 only becomes substantial but also can change sign. In
 Fig.~\ref{noise1} we plot the shot noise cross-correlations as
 function of the gate voltage, which tunes the strength of the left
 insulator in the system. As the effective barrier strength changes,
 one sees negative cross-correlations turning positive for $L < \xi$.
 This indicates that a gate voltage can tune the entanglement
 properties. More interesting is the case $L=\xi$, where noise
 cross-correlations turn completely positive enabled by the strong CAR
 signal. In the specular regime the noise is enhanced. This can be
 understood from Fig.~\ref{car-ec} where the CAR signal in the
 specular regime is double than that of the normal case. The behavior
 depicted in Fig.~\ref{noise1} is of significance for the experimental
 detection of entanglement in solid state systems. It shows that a
 gate voltage can change the sign of noise cross-correlations unlike
 that predicted for normal metal counterparts. It is worth mentioning
 that for $L\gg\xi$ the magnitude of the noise cross-correlations are
 very much reduced (not plotted) but one can also see completely
 positive noise cross-correlations.

\section{Conclusions}

Recent CAR experiments~\cite{Beckmann} are the next generation in
detecting the splitting of Cooper pairs into different leads, thus
probing entanglement in the context of nanophysics. In this work we
provide the results of noise cross-correlation spectra as a function
of gate voltage for a NISIN graphene junction. The Fano factor (not
presented here) is also on predictable lines and shows a spike in case
of enhanced positive noise cross-correlations, indicating bunching. We
point out the novel phenomena of specular crossed Andreev reflection,
which enhances noise cross-correlations. The settings envisaged in
this work are experimentally accessible. A typical s-wave
superconductor like Aluminium has a coherence length of $\xi=1600 nm$.
Since the proximity effect induces superconducting correlations in
graphene, an Aluminium superconductor on top of the graphene layer
would give rise to a similar correlation length. This separation would
not be a challenge since crossed Andreev reflection measurements are
carried out routinely at less than these lengths. Further, the
superconducting gap in Aluminium is $1meV$, while the typical Fermi
energy in normal doped graphene is around $80 meV$. In our study we
have considered for certain situations $E_{F}=100\Delta$, i.e., $E_{F}
\gg \Delta $, which corresponds to undoped graphene, while $E_{F} \ll
\Delta$ can be tuned via doping graphene. These values are realistic
and thus obviate any reasons for scepticism. Employing these entangled
states for quantum information processing will increase the allure of
graphene.
\section{Acknowledgments}
We would like to thank Chris Marrows and Graham Creeth for stimulating discussions. This work was supported by EPSRC, the EU grants EMALI and SCALA, and the Royal Society.


\begin{thebibliography}{99}

\bibitem{Martin} G. B. Lesovik, T. Martin and G. Blatter,
Eur. Phys. J. B \textbf{24}, 287 (2001).
\bibitem{ref-on-noise} M. P. Anantram and S. Datta, Phys. Rev. B {\bf 53},
16390 (1996).
\bibitem{pt-noise} C. W. J. Beenakker and C. Schonenberger,
Physics Today  May 2003, page 37.
\bibitem{melin-cb} R. Melin, C. Benjamin and T. Martin,
\prb {\bf 77}, 094512 (2008).
\bibitem{buttiker} P. Samuelsson and M. B\"{u}ttiker,  J. of Low Temp. Phys. {\bf 146}, 115 (2007).
\bibitem{loss-entangle} P. Recher, E. V. Sukhorukov and D. Loss,
Phys. Rev. B {\bf 63}, 165314 (2001).
\bibitem{kontos} B. R. Choi, et. al., \prb {\bf 72}, 024501 (2005).

\bibitem{heersche} H. B. Heersche, et. al., Nature {\bf 446}, 56 (2007).
\bibitem{teleport} C. W. J. Beenakker and M. Kindermann, Phys.
Rev. Lett. {\bf 92}, 056801 (2004).
\bibitem{graphene-rmp} C. W. J. Beenakker, arxiv:0710.3848.
\bibitem{graphene-sudbo} J. Linder and A. Sudbo, \prl {\bf 99}, 147001 (2007); arxiv: 0712.083.
\bibitem{Duhot-Melin} S. Duhot and R. M\'elin, Eur. Phys. J. B {\bf 53}, 257 (2006); G. Deutscher and D. Feinberg, \apl {\bf 76}, 487 (2000);  J. M. Byers and  M. E. Flatte, \prl {\bf 74}, 306 (1995); Colin Benjamin, \prb {\bf 74}, 180503(R) (2006).

\bibitem{been-gra-supercon} C. W. J. Beenakker, \prl {\bf 97},
067007 (2006); arxiv:0710.3848.
\bibitem{cayssol} J. Cayssol, \prl {\bf 100}, 147001 (2008).

\bibitem{sengupt-gra} S. Bhattacharjee and K. Sengupta, \prl {\bf
97}, 217001 (2006).
\bibitem{Beckmann} D. Beckmann, H.B. Weber and H.v. L\"ohneysen,
 Phys. Rev. Lett. {\bf 93}, 197003 (2004); S. Russo, M. Kroug, T.M. Klapwijk and A.F. Morpurgo,
 Phys. Rev. Lett. {\bf 95},  027002 (2005); P. Cadden-Zimansky and V. Chandrasekhar,
Phys. Rev. Lett. {\bf 97}, 237003 (2006).
\end{thebibliography}
\end{document}